# Multiple strain-induced phase transitions in LaNiO$_3$ thin films

M.C. Weber[1,2], M. Guennou[1],
N. Dix[3], D. Pesquera[3], F. Sánchez[3], G. Herranz[3], J. Fontcuberta[3],
L. López-Conesa[4], S. Estradé[4], F. Peiró[4],
Jorge Iñiguez[1,3] & J. Kreisel[1,2,*]

[1] Materials Research and Technology Department, Luxembourg Institute of Science and Technology, 41 rue du Brill, L-4422 Belvaux, Luxembourg

[2] Physics and Materials Science Research Unit, University of Luxembourg, 41 Rue du Brill, L-4422 Belvaux, Luxembourg

[3] Institut de Ciència de Materials de Barcelona (ICMAB-CSIC), Campus UAB, Bellaterra 08193, Catalonia, Spain

[4] Laboratory of Electron Nanoscopy (LENS-UB), Electronics Department, University of Barcelona, c/ Martí i Franquès 1, Barcelona 08028, Catalonia, Spain

* Corresponding author:   jens.kreisel@list.lu

**Abstract**

Strain effects on epitaxial thin films of LaNiO$_3$ grown on different single crystalline substrates are studied by Raman scattering and first-principles simulation. New Raman modes, not present in bulk or fully-relaxed films, appear under both compressive and tensile strains, indicating symmetry reductions. Interestingly, the Raman spectra and the underlying crystal symmetry for tensile and compressively strained films are different. Extensive mapping of LaNiO$_3$ phase stability is addressed by simulations, showing that a variety of crystalline phases are indeed stabilized under strain which may impact the electronic orbital hierarchy. The calculated Raman frequencies reproduce the principal features of the experimental spectra, supporting the validity of the multiple strain-driven structural transitions predicted by the simulations.

1. Introduction

In recent years, the engineering of electron orbital occupation in transition-metal perovskite oxides with *AMO$_3$* composition –where *A* usually stands for an alkali or a rare earth and *M* for a transition metal– has been a powerful tool to finely tune the interactions between charge, spins, and phonons, leading even to the discovery of emergent properties. Pivotal to this progress is the use of thin film heteroepitaxies grown on different substrates. The structural mismatch between film and substrates imposes structural constraints via the necessary connectivity of the metal coordination polyhedral of the substrates and the films.  In turn, these constrains allow tailoring the film structure[1]. Substrate-induced strain is the toggle switch that permits imposing elastic stress, compressive or tensile, on films, and it has been much used to modulate properties of a variety of perovskite oxides.

In this context, LaNiO$_3$ (LNO) films and LNO-based heterostructures have recently received much attention. The interest is mainly motivated by a strong similarity of low-spin Ni$^{3+}$ ions, having an electronic configuration 3d$^7$: t$_{2g}^6$e$_g^1$,  to Cu$^{2+}$ ions in the high-temperature superconducting cuprates (HTSCs, with 3d$^9$: $t_{2g}^6$ $e_g^3$). In the HTSC materials, due to their layered structure, the $z^2$ orbitals are pushed down in energy and are fully occupied, whereas conductivity is confined to the partially occupied $x^2$-$y^2$ orbitals. Claims have been made that, nickelate films would mimic HTSCs if under appropriate (tensile) strain the orbital degeneracy ($z^2$, $x^2$-$y^2$) of the $e_g$ manifold in LNO could be broken and the $x^2$-$y^2$ pushed down in energy [2]. Subsequently, the growth and properties of LNO films on different substrates have been vastly explored.

When the film is subjected to a compressive or tensile stress imposed by substrates, the LNO lattice may either deform elastically in a continuous way, or switch to a different crystallographic structure. These strain-induced structural transformations are central to understand some dramatic changes of physical properties such as metal-insulator transitions observed on LNO thin films and LNO multilayers[3–6], or to rationalize reported spectroscopic results. For instance, Chakhalian et al.[7], Freeland et al.[8], and Tung et al.[9] reported X-ray absorption results at Ni edges of LNO films grown on LaAlO$_3$ (LAO) and SrTiO$_3$ (STO) substrates, which impose compressive and tensile stresses, respectively, supported by first-principles Density Functional Theory (DFT) calculations. It was concluded that compressive and tensile strain on ultrathin layers of LNO have very different effects: while tensile strain was found to induce a Jahn-Teller like distortion of the NiO$_6$ coordination octahedra, compressive strain yielded two distinguishable (different volume) regular NiO$_6$ octahedra, somehow suggesting Ni$^{(3\pm\delta)}$ charge disproportionation. May et al.[10] used diffraction techniques and high-flux synchrotron X-ray beams to explore the structure of ultrathin LNO films grown on LAO and STO substrates. In order to determine the actual crystallographic structure of the films, the diffracted intensities were measured and compared to those obtained by calculation of suitable structure factors. However, the data analysis of May et al.[10] is complicated by the presence of twins in the films, which originate either from the film itself or from pre-existing twins in the substrates. Yet, it was concluded that, in thin film form, LNO changes its symmetry from the bulk rhombohedral (*R-3c*) space group to a monoclinic *C2/c* group. Interestingly, the same symmetry group was proposed under both compressive and tensile stresses. DFT calculations were also used to explore strain effects on LNO films. It was found that *C2/c* was the most stable phase, irrespective of strain sign, and in spite of the proximity of a first-order isosymmetric phase transition close to zero strain. Notably, in this monoclinic *C2/c* phase, the Ni$^{3+}$ ions occupy a single Wyckoff position, which is not compatible with the previous proposition of two non-equivalent NiO$_6$ octahedra under compression.

The difficulties to characterize the LNO films are partly due to the fact that X-ray diffraction techniques give an average structure that reflects the coherently diffracting part of the film and requires a substantial amount of matter to obtain intensities suitable for a quantitative analysis of subtle structural distortions. This is particularly challenging in thin films where the sample volume is very small, and twinning pervasive.

Here we tackle the problem of structural analysis of LNO ultrathin films by using Raman scattering, known to be a well-adapted technique for the observation of strain effects[11–14] and structural transitions in thin films[13,15,16], in combination with X-ray diffraction, electron microscopy and DFT simulations. LNO films of various thicknesses have been grown on substrates imposing either compressive (LAO) or tensile stress ((LaAlO$_3$)$_{0.3}$(Sr$_2$TaAlO$_6$)$_{0.7}$, LSAT) on the films. The pseudocubic bulk cell parameter of LNO is 3.838 Å[17] and that of LAO and LSAT is 3.795 Å and 3.865 Å, respectively, corresponding to a structural mismatch of 1.1% (compressive, LAO) and -0.5% (tensile, LSAT). Structural analyses have been performed by X-ray diffraction and scanning transmission electron microscopy, in order to check sample quality and carry out a preliminary investigation of strain effects.

Within the experimental accuracy of these techniques, no discernible changes of film symmetry are found in strained films. Next, Raman scattering experiments have been performed, and dramatic and well discernible changes in the pattern have been observed, depending on the strain sign and magnitude, which are taken as fingerprints of symmetry changes. Finally, DFT calculations have been run aiming at resolving structure stability under strain. A series of phase transformations with associated symmetry changes are identified. The DFT results for the Raman modes of the theoretically-identified phases reproduce the basic features of the measured Raman spectra, and allow rationalization of the changes observed experimentally. Further, this agreement supports the accuracy of the non-trivial structural transformations predicted by the simulations.

2. **Experimental**

LNO films were epitaxially grown by pulsed laser deposition on (001) oriented LAO and LSAT single crystal substrates, using a KrF excimer laser ($\lambda$ = 248 nm) with a fluence of around 1.5 J cm$^{-2}$ and a repetition rate of 5 Hz. Deposition on both substrates was performed simultaneously at an oxygen pressure of P = 0.15 mbar and keeping the substrates at a temperature of T = 700 °C. The number of laser pulses was varied to obtain films with nominal thickness of 14 nm, 35 nm, and 130 nm, according to previous growth-rate calibrations.

X-ray diffraction (XRD) characterization was done by measuring $\theta$-$2\theta$ scans and $\omega$-scans, as well as reciprocal space maps around asymmetric reflections; this was done using Cu-K$\alpha$ radiation in a Siemens D500 and a PANalytical X'pert diffractometer, respectively. Grazing incidence measurements were used in some experiments to determine the in-plane cell parameters. Raman spectra were recorded with a Renishaw inVia Reflex Raman Microscope. Experiments were conducted in micro-Raman mode at room temperature, using a red 633 nm line from a HeNe laser as exciting wavelength. As transition metal oxides are often subjected to overheating by the laser, it was carefully checked that the used laser power was low enough to avoid any modification of the spectral signature. The reproducibility of spectra on different places of the sample was also verified. Samples were prepared for scanning transmission electron microscopy (STEM) observation by mechanical flat polishing and subsequent low-angle argon ion milling up to electrotransparency. High-angle annular dark-field imaging (HAADF) images were obtained in a probe-corrected FEI Titan microscope operated at 300 kV and equipped with a high brightness X-FEG field emission electron gun.

3. **X-ray Diffraction**

Wide angular range $\theta$-$2\theta$ scans of LNO films on both substrates displayed only (00*l*) reflections of the LNO films without indications of secondary crystalline phases. In Figs. 1a and 1b we present a zoom of the $\theta$-$2\theta$ scans, around the (002) reflection, of the 14 nm thick LNO films on LAO and LSAT substrates, respectively. Intensity profiles have been analysed using dynamical diffraction theory[18] and used to extract the out-of-plane cell parameters (c-axis). The pattern of the LNO/LAO film (Fig. 1a) shows that the LNO (002) reflection occurs at a smaller angle than the (002) reflection of the LAO indicating that the c-axis parameter of the latter (c(LNO/LAO) ≈ 3.91 Å) is larger than that of substrate (3.795 Å). This c(LNO/LAO) value is clearly larger than the corresponding bulk value of 3.838 Å[10]. This reflects the in-plane compressive stress imposed by the LAO substrate on LNO. In LNO/LSAT (Fig. 1b) the situation is reversed: the out-of-plane lattice parameter (c(LNO/LSAT) = 3.81 Å) is smaller than in bulk one due to the tensile stress imposed by LSAT (3.86 Å) on LNO.

The out-of-plane cell parameters were similarly determined for films of all thicknesses. Reciprocal space maps (RSM) (some of them are shown in Fig. 2) and grazing incidence diffraction patterns were used to quantify in-plane cell parameters of the thicker and thinnest films, respectively (Table I). It can be appreciated in Table I that the c-axis of LNO gradually evolves towards its bulk value when the film thickness increases, indicating a gradual strain relaxation.

TABLE I. Cell parameters (c, a) of the LaNiO$_3$ thin films of different thicknesses (130, 35, and 14 nm) and the corresponding strain ($\varepsilon$). The differences between the out-of-plane cell parameters of the 14 nm films and the substrate ($\varepsilon_{subs}$) are included.

|  |  | 130 nm | $\varepsilon$ (%) | 35 nm | $\varepsilon$ (%) | 14 nm | $\varepsilon$ (%) | $\varepsilon_{subs}$ (%) |
|---|---|---|---|---|---|---|---|---|
| LNO/LAO | c (Å) | 3.85 | 0.31 | 3.87 | 0.83 | 3.91 | 1.88 | 3.03 |
| LNO/LAO | a (Å) | 3.81 | -0.73 | 3.79 | -1.25 | 3.81 | -0.73 |  |
|  |  |  |  |  |  |  |  |  |
| LNO/LSAT | c (Å) | 3.82 | -0.47 | 3.81 | -0.73 | 3.81 | -0.73 | -1.30 |
| LNO/LSAT | a (Å) | 3.84 | 0.05 | 3.87 | 0.83 | 3.86 | 0.57 |  |

In Figs. 1c and 1d we show $\omega$-scans around (002) for LNO/LAO and LNO/LSAT films, respectively. It can be appreciated that the rocking curve of the thinnest films presents a single peak. However, when increasing film thickness, and irrespective of the substrate, the reflections progressively broaden and eventually two discernible peaks, symmetric around $\omega$ = 0, are well visible for the thickest (130 nm) films. These features may be signatures of twinning. As twinning is not expected in fully (00l)-textured tetrahedrally distorted structures, it may result from a strain-induced lowering of the film symmetry. We notice that twinning is not visible in the thinnest LNO/LAO films suggesting that the observed twinning does not arise from the LAO substrate, but is a film property.

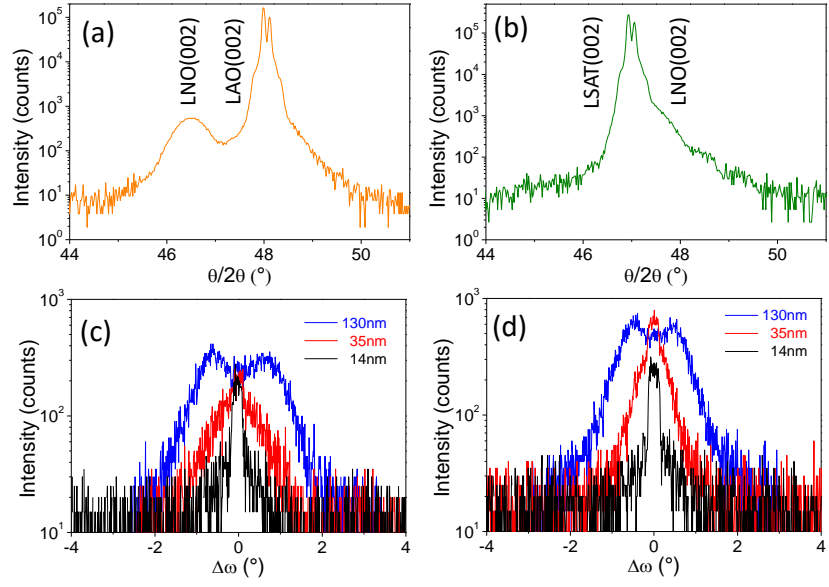

FIG. 1. X-ray diffraction θ-2θ scans around the (002) reflections of 14 nm thick LNO films grown on (a) LAO and (b) LSAT substrates, respectively, and fittings to XRD patterns (continuous black lines). The ω–scans around the (002) reflection of LNO films of different thicknesses are plotted in (c) LAO and (d) LSAT, respectively.

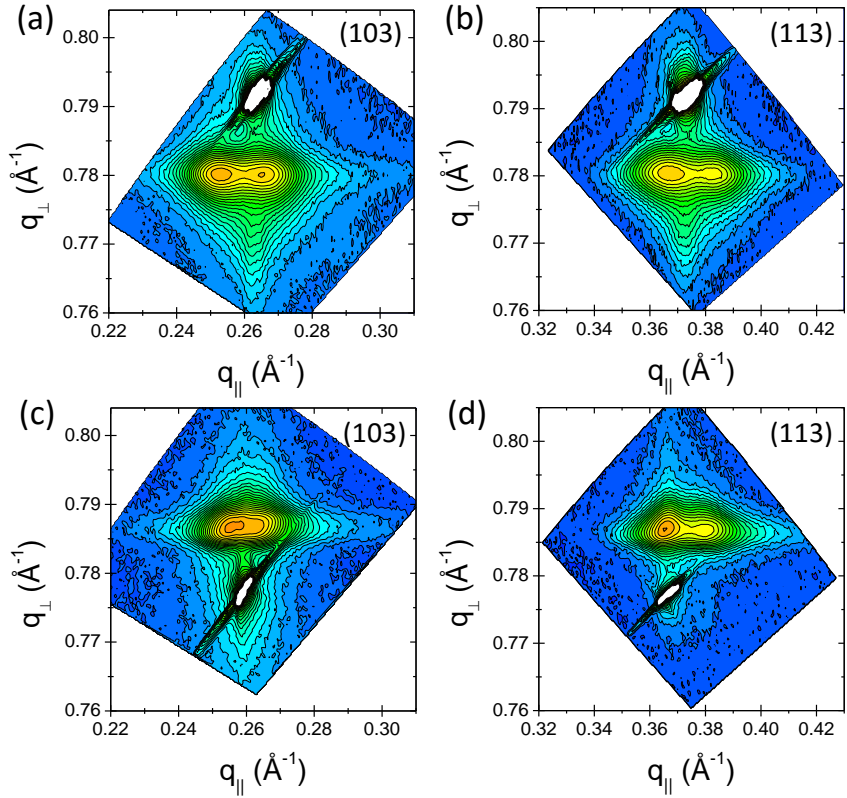

FIG. 2. X-ray diffraction reciprocal space maps around the (103) and (113) reflections of 130 nm thick LNO films grown on (a, b) LAO and (c, d) LSAT, respectively.

In Fig. 2(a-d) we show illustrative examples of the RSMs collected around the (103) and (113) reflections. In the RSM of the compressively strained film (LNO(130 nm)/LAO) in Figs. 2(a) and 2(b), it can be appreciated that the $q_z$ values are smaller than those of the substrate (brightest spots), and a double-peak associated with the film emerges. This is compatible with either the presence of twinning or a monoclinic distortion, which are both consistent with the ω-scans presented above. These features can also be identified in the 35 nm films. For the 14 nm film, the weak intensity of the LNO reflections in the RSM makes it impossible to assess if the cell symmetry is preserved in the thinnest film.

   **4. High Angle Annular Dark Field imaging**

Figure 3 shows HAADF images of the 14 nm LNO films grown on a) LAO and b) LSAT substrates, collected along the [001] zone axis. The interface between substrate and layer is atomically sharp, reflecting the high quality of the films. The fast Fourier transforms (FFTs – top left insets in the images) show a coherent epitaxy between the films and substrates. Superimposed color maps correspond to the in-plane and out-of-plane strains obtained from geometric phase analysis (GPA), considering the substrate as the reference[19]. It can be appreciated that in both cases the LNO films grow fully coherently with the substrate, while the out-of-plane cell parameter changes according to the sign of the in-plane strain induced by the substrate, that is, expanding c-axis for the compressive LAO and shrinking it for the tensile LSAT. More precisely, the LNO/LAO film (Fig. 3a) displays an out-of-plane expansion of about +2% whereas the LNO/LSAT film (Fig. 3b) displays a contraction of about -1%. These values are in rough agreement with those obtained from the XRD (+3.03 % and -0.7%, respectively; Table I), the differences could be attributed to the possible occurrence of strain gradients in the films. Thicker films of 35 nm LNO grown on LAO and LSAT also present atomically sharp substrate/film interfaces, with equivalent strain states in the first few nanometers along the growth direction, as found by GPA.

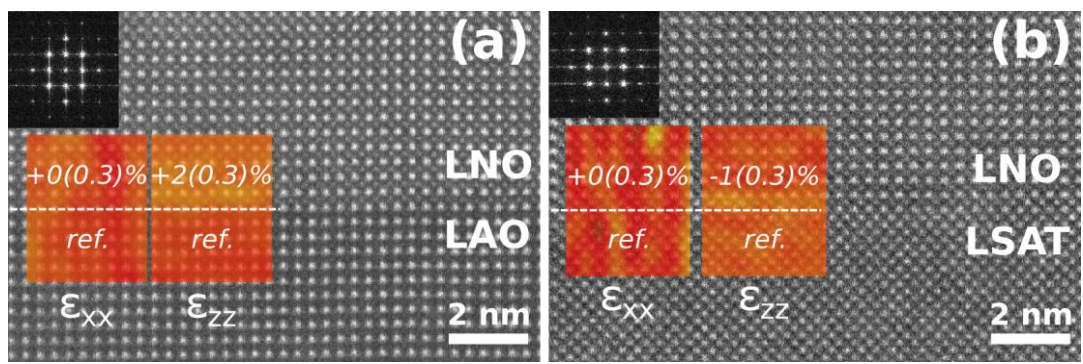

FIG. 3. HAADF cross section image and corresponding FFT (top left insets) of the 14 nm LNO film grown on: a) LAO and b) LSAT. Color maps correspond to the in-plane ($\varepsilon_{xx}$) and out-of-plane ($\varepsilon_{zz}$) components of the strain tensor obtained from GPA with the strain values in % and in brackets the corresponding error.

**5. Raman scattering**

In order to get more insight into the effect of strain on the film's structure, we investigated the LNO films by Raman scattering. According to group theory, the rhombohedral R-3c bulk crystal structure of LNO gives rise to five Raman-active modes $\Gamma_{Raman}$ = $A_{1g}$ + 4 $E_g$. The different lattice normal modes have been calculated by Gou et al.[20] and are in agreement with earlier published experimental Raman spectra[21].

Figure 4 presents Raman spectra for seven films: Three of them under compressive strain (14 nm, 35 nm, and 130 nm, LNO/LAO), three under tensile strain (14nm, 35nm, and 130 nm, LNO/LSAT), and, for comparison, a spectrum of a strain-free polycrystalline LNO film on a Si substrate (taken from ref. 21). All Raman spectra are well defined, remarkably even for the very thin 14 nm films. As expected, we observe for the thinnest films a superposition of spectra from the film and the respective substrate, while the spectra from the much thicker 130 nm films show no background signal from the substrate. The Raman spectra for all films on LSAT and LAO display spectral signatures that differ from that of the unstrained film, both in terms of band position and number of bands. We first discuss the effect of tensile strain of LNO/LSAT. The spectral signature for the 35 nm and 14 nm films (red spectra) are remarkably similar and show, instead of the $E_g$ peak at 400 cm$^{-1}$ of bulk LAO, a broad and multiple band that can be deconvoluted in three components at about 373, 393 and 409 cm$^{-1}$. This clearly points to a strain-induced phase transition to a new space group different from the bulk rhombohedral R-3c structure. The increase in the number of bands suggests a symmetry lowering involving not only the splitting of the degenerated $E_g$ modes but also the emergence of new Raman active modes. For the thicker 130 nm film, the spectrum (in blue) is different from both thinner films and bulk samples. It also shows clear signs of a symmetry lowering, especially through the noticeable splitting of the two $E_g$ bands around 150 and 400 cm$^{-1}$. This suggests the presence of a third and again different structure, although it is difficult in this case to conclusively rule out the possibility of a phase mixture involving the R-3c phase, which might be stabilized by partial strain relaxation. The region II, represented by the $A_g$ mode, shows no significant changes.

We next discuss the spectra of compressively strained LNO/LAO films. We also observe new spectral signatures that differ from those observed under no or tensile strain. Most remarkably, we find a splitting of the 150 cm$^{-1}$ $E_g$ mode (Region I), a weak but distinct shoulder at 390 cm$^{-1}$ (Region III), while the band around 214 cm$^{-1}$ (Region II) remains a singlet, as already observed in the tensile LNO/LSAT case above. This is in agreement with Hepting et al.[22] who reported the presence of a band at 390 cm$^{-1}$ in LNO/LAO films; the splitting of the 150 cm$^{-1}$ was not observed in their case probably due to the limited frequency range scanned. Again, the increasing number of bands suggests a symmetry lowering, but the distinct signature demonstrates that the actual structure is different in tensile and compressively strained films. Although the splitting around 150 cm$^{-1}$ in region I is masked for the 14 nm and 35 nm films, the remaining spectral signatures suggest that all investigated films on LAO present the same overall features.

After the qualitative observation of the different spectral signatures, it is interesting to consider the individual modes in more detail to discuss the observed changes. According to an earlier experimental[21] and theoretical[20] assignment, the vibrational pattern of the $E_g$ mode at 156 cm$^{-1}$ in rhombohedral LNO corresponds to pure La vibrations in the hexagonal (001) plane along the *a* and *b*-axis. It splits in both the tensile (not so visible in the 35 nm and 14 nm cases) and compressive state, indicating a symmetry lowering for the La vibrations. On the other hand, the $E_g$ mode at 400 cm$^{-1}$ (region III) is attributed to a bending deformation of the oxygen NiO$_6$ octahedra[20,21]. This region III probes the internal distortion of the octahedra, and the splitting of the $E_g$ mode, particularly marked under tensile strain, suggests the presence of distinct Ni-O bond-lengths inside the NiO$_6$ octahedra. Interestingly, the $A_g$ mode at 214 cm$^{-1}$ (region II) shows no change in frequency throughout the whole series. This mode represents the [111]-octahedral tilting vibration.[20] Thus, we deduce that the strain induced by the present substrates is mediated rather through deformation of octahedra than through changes of the octahedral tilt angle.

In summary, Raman spectroscopy gives evidence for significant strain-induced structural changes and polymorphism in LNO thin films, with at least two structures markedly different from that of bulk LNO. In order to test our above interpretations, and especially to elucidate the actual symmetry of the strain-induced phases, we conducted a first-principles investigation.

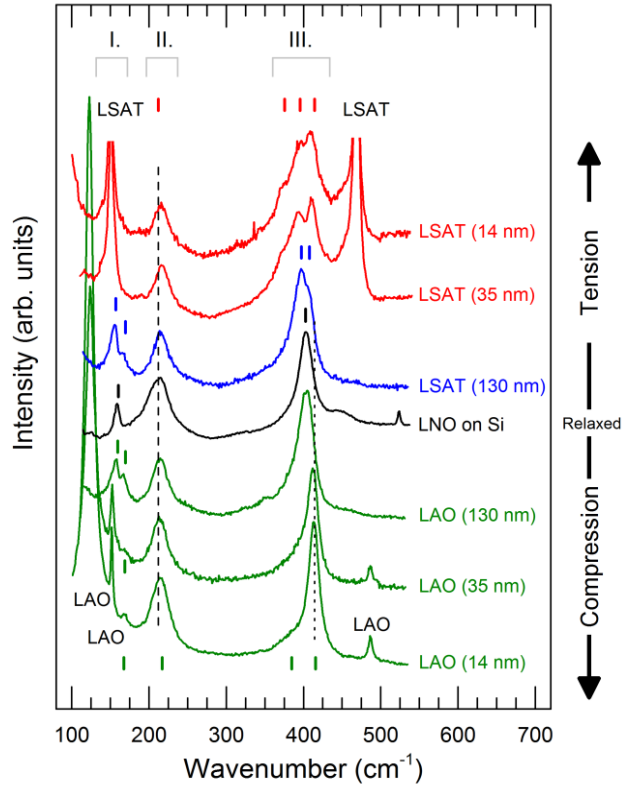

FIG. 4. Raman spectra of LaNiO$_3$ of different indicated thicknesses either on LAO or LSAT substrates. The ticks and lines allow following the band positions and the letters I, II and III in the top indicate the three spectral regions discussed in the text.

**6. First-principles calculations**

For the calculations we used DFT[23] within the local density approximation (LDA)[24] as implemented in the VASP package[25,26]. The use of the LDA without any sort of strong-correlation correction was adopted as it has been shown to provide a qualitatively correct representation of LNO's bulk metallic and paramagnetic ground state[27]. We used the so-called projector augmented method to represent the ionic cores[26,28], treating explicitly the electrons in the following atomic orbitals: Ni's 3p, 3d, and 4s; La's 5s, 5p, 5d, and 6s; and O's 2s and 2p. We represented the electronic wave functions in a basis of plane waves cut off at 500 eV. Most of our calculations were done in a 20-atom cell that is essentially given by the lattice vectors b$_1$ = (a$_1$+a$_2$)/2$^{1/2}$, b$_2$=(-a$_1$+a$_2$)/2$^{1/2}$, and b$_3$=2a$_3$, where a$_1$ =

a(1,0,0), a$_2$ = a(0,1,0), and a$_3$ = a(0,0,1) define the elemental 5-atom perovskite unit expressed in a Cartesian basis. Note that this cell is compatible with the rhombohedral structure (R-3c space group) of the ground state of bulk LNO, the structures proposed in the literature for this kind of thin films[10], and other typical combinations of the O$_6$-octahedra rotations characteristic of perovskite oxides. We considered a Γ-centered 5x5x5 k-point grid for calculating integrals in the first Brillouin Zone associated to this cell, which proved to be sufficiently well converged for the current purposes (e.g., phonon frequencies obtained for denser grids typically differ by 1 cm$^{-1}$ or less).

The epitaxial constraint corresponding to a squared (001)-oriented substrate was imposed by fixing the b$_1$ and b$_2$ vectors during the structural relaxations, with the value of the parameter a being equal to the substrate lattice constant a$_{sub}$. All the other lattice and atomic degrees of freedom were relaxed during the structural optimizations, which were typically run under no symmetry restrictions (so that distortions of the initially-expected structures could be explored) and stopped when the residual forces on the atoms were below 0.01 eV/A$^2$. Phonons were obtained from the corresponding force-constant matrices, which were computed by finite differences. A standard symmetry analysis[29] allowed us to identify the Raman-active modes, for which we report their frequencies. Note that obtaining Raman intensities in a metal like LNO remains non-standard from the first-principles perspective, and we did not attempt such a calculation in this work.

It should be stressed that, as it is customarily done in this type of DFT investigations, we used a minimal simulation box that is periodically repeated, neglecting some effects (twinning, existence of surfaces and interfaces) that are present in the real films. In fact, the elastic constraint, which is chosen to mimic the case of an ideal fully-strained mono-domain case, is the only ingredient in the simulations that models the specific conditions of a thin film.

**6.1 Stable phases as a function of epitaxial strain**

We investigated the phase diagram of epitaxial LNO films on a (001)-oriented substrate covering a wide range of values for the substrate lattice parameter a$_{sub}$. More precisely, we moved from strongly compressive substrates (well below LAO, which displays a$_{sub}$ ≈ 3.79 Å (a√2 = 5.36 Å)) to strongly tensile substrates (up to STO with a$_{sub}$ ≈ 3.90 Å (a√2 = 5.52 Å)). For the initial exploratory runs, we considered an epitaxially deformed version of bulk LNO's rhombohedral ground state, as well as the structures experimentally found in ref. 10 for films grown on LAO and STO. In all cases we introduced some small symmetry-breaking distortions to facilitate the structural search for energy minima. This allowed us to identify some new, previously unreported structures that we predict become stable in certain epitaxial strain ranges.

Our main results are summarized in Fig. 5, which shows the strain dependence of the four most stable structures identified in our calculations. A sequence of transitions as a function of a$_{sub}$ is clearly observed. Further, Fig. 5 also shows the basic structural features of these phases, i.e., their anti-phase O$_6$-rotational patterns about the three pseudo-cubic axes of the perovskite lattice (central panel) as well as the distances between the Ni atoms and the surrounding oxygens in the corresponding O$_6$ octahedron (bottom panel).

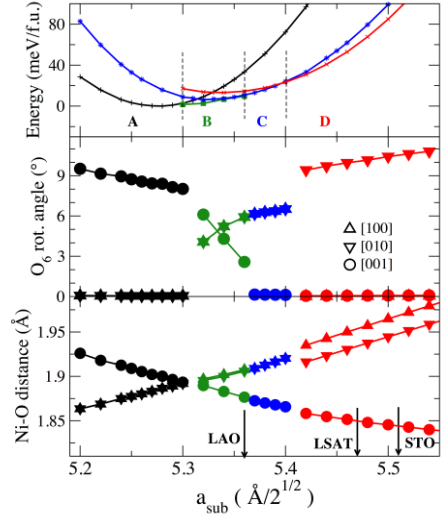

FIG. 5. Computed energy (top panel), rotation angles of $O_6$ octahedra (central), and Ni—O distances (bottom) as a function of the substrate lattice constant $a_{sub}$. We identified four stable phases (see text) indicated in the top panel: Phase A has space group *I4/mcm*; phase B: *C2/c*; phase C: *Imma*; phase D: *Fmmm*. In all the found phases there is a single crystallographic position for Ni (i.e., all $NiO_6$ groups are equivalent by symmetry). The rotations in the central panel correspond to anti-phase patterns, with all the $O_6$ octahedra in the cell tilting by the indicated angles. Rotation amplitudes about the three pseudo-cubic axes are shown. In the bottom panel, the Ni—O distances for bonds aligned with the three pseudo-cubic directions are indicated. The values of $a_{sub}$ corresponding to the experimental LAO, LSAT, and STO substrates are indicated with arrows. Note that LDA has a well-known tendency to overbind, i.e., to render equilibrium lattice constants that are smaller than the experimentally measured ones by about 1 %. Hence, in the figure, the LDA positions of the reference substrates would be shifted by about -0.05 Å with respect to the shown experimental ones.

In Fig. 6 we sketch the identified phases and indicate their space groups, as deduced from our relaxed structures by employing the crystallographic tools in the Materials Studio package.[30] Full structural details of representative cases are given in Table II.

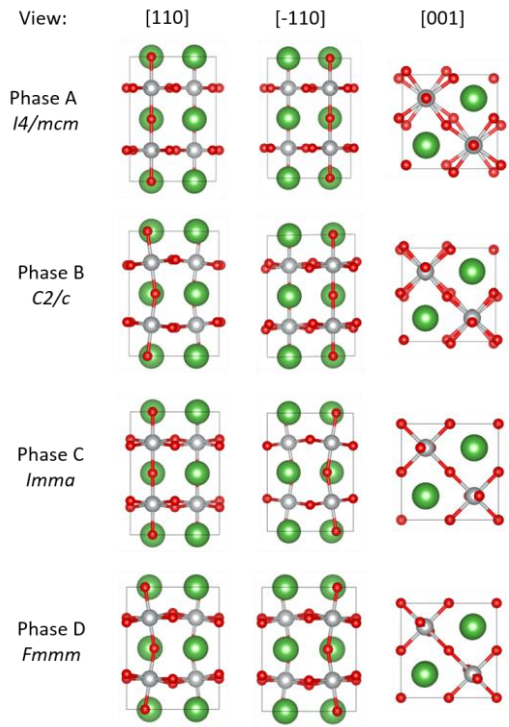

FIG. 6. Phases identified from first-principles simulations, as viewed along three different directions, indicated in the pseudo-cubic setting. Note that Phase B can be associated to the *C2/c* space group or, equivalently, to *I2/a*, as indicated in Table II. Likewise, Phase C can be associated to the *Imma* space group or, equivalently, to *Ibmm*.

TABLE II. Representative structures for phase A (at $a_{sub}$ = 5.25/√2 Å), phase B (at $a_{sub}$ = 5.32/√2 Å), phase C (at $a_{sub}$ = 5.36/√2 Å) and phase D (at $a_{sub}$ = 5.47/√2 Å), as obtained from the elastically constrained structural relaxations described in the text. We have adopted settings that are compatible with the orientation of our structures as simulated on a square (001)-oriented substrate; for phases B and C, these settings do not coincide with the (equivalent) ones most commonly mentioned in the literature when discussing similar structures, which are indicated for clarity. Note that for phase D we give the structure in a conventional cell that is twice as big as the one used in the simulations (as indicated by the √2 factors for the *a* and *b* lattice constants). For reference, we also give the relaxed structure of bulk LaNiO$_3$ (*R-3c* space group) in a hexagonal setting.

| *I4/mcm* | | a = b = 5.250 Å    c = 7.636 Å | | |
| --- | --- | --- | --- | --- |
| | | $\alpha = \beta = \gamma = 90°$ | | |
| Atom | Wyc. | x | y | z |
| La | 4b | 0 | 1/2 | 1/4 |
| Ni | 4c | 1/2 | 1/2 | 0 |
| O | 4a | 0 | 0 | 1/4 |
| O | 8h | 0.2120 | 0.2880 | 0 |
| *I2/a* | | a = b = 5.330 Å    c = 7.484 Å | | |
| (equiv. *C2/c*) | | $\alpha = \gamma = 90°$    $\beta = 91.006°$ | | |
| Atom | Wyc. | x | y | z |
| La | 4e | 1/2 | 0.9994 | 1/4 |
| Ni | 4a | 0 | 0 | 0 |
| O | 4e | 1/2 | 0.4563 | 1/4 |
| O | 8f | 0.7312 | 0.7691 | 0.0222 |
| *Ibmm* | | a = b = 5.360 Å    c = 7.414 Å | | |
| (equiv. *Imma*) | | $\alpha = \beta = \gamma = 90°$ | | |
| Atom | Wyc. | x | y | z |
| La | 4e | 0.0014 | 0 | 3/4 |
| Ni | 4b | 0 | 1/2 | 0 |
| O | 4e | 0.5524 | 0 | 3/4 |
| O | 8g | 3/4 | 3/4 | 0.0266 |
| *Fmmm* | | a = b = √2×5.470 Å    c = 7.232 Å | | |
| | | $\alpha = \beta = \gamma = 90°$ | | |
| Atom | Wyc. | x | y | z |
| La | 8g | 0.7465 | 0 | 1/2 |
| Ni | 8c | 1/2 | 1/4 | 1/4 |
| O | 8f | 1/4 | 1/4 | 1/4 |
| O | 8i | 0 | 0 | 0.7114 |
| O | 8h | 1/2 | 0.2997 | 0 |
| *R-3c* | | a = b = 5.346 Å    c = 12.834 Å | | |
| | | $\alpha = \beta = 90°$    $\gamma = 120°$ | | |
| Atom | Wyc. | x | Y | z |
| La | 6b | 0 | 0 | 0 |
| Ni | 6a | 1/3 | 2/3 | 5/12 |
| O | 18e | 0.5433 | 0.5433 | 0 |

The DFT calculations predict that, for substrates with small $a_{sub}$, LNO presents a structure with *I4/mcm* space group and with the rotational pattern labelled $a^0a^0c^-$ in Glazer's notation.[31] For brevity, we refer to this as "phase A" in the following. This is a natural finding as such a rotational pattern is known to be effective in accommodating the in-plane compression.

As $a_{sub}$ increases, the LNO films undergo an abrupt first-order transition to a "phase B" with *C2/c* space group in which the in-plane tilts are recovered; the new structure can thus be described as $a^-a^-c^-$. This structure is the epitaxially-modified version of bulk LNO that one would expect to observe in a (001)-oriented film of this material by default and, indeed, our calculations indicate that it is stable for $a_{sub}$ values close to the nominal best match with the computed bulk LNO lattice constant (3.780 Å = 5.346 / √2 Å; see Table II). Interestingly, our obtained structure (Table II) does not coincide exactly with the one reported by May *et al.*[10] for LNO on LAO substrates. Beyond certain differences in the DFT methods employed, we believe this disagreement may be partly attributed to the fact that, in this strain range, we obtained many $a^-a^-c^-$-like structures that are local minima of the energy; here we are reporting the solution that is most stable according to our calculations.

As we continue to increase $a_{sub}$, we get a continuous transformation to a new structure, with *Imma* space group, whose essential feature is the disappearance of the out-of-plane octahedral rotations ("phase C"). This is a natural way in which perovskites can accommodate a tensile in-plane strain and, hence, the occurrence of such a structure is not surprising. This is the structure that May *et al.*[10] reported for the LNO on STO substrates.

Finally, and surprisingly, as we keep increasing $a_{sub}$, we find a strongly-discontinuous transition to a new "phase D" with the *Fmmm* space group and a peculiar $a^-b^0c^0$ rotational pattern. As shown in Fig. 5(a) this structure is very close in energy to the *Imma* solution, but nevertheless becomes the ground state for $a_{sub} > 3.83$ Å according to our calculations. The peculiarity of the rotational pattern is that this structure presents tilts only about one of the in-plane pseudo-cubic axes; as a result, the in-plane directions $a_1$ and $a_2$ are not symmetric, in contrast with what is usually assumed for this kind of films. Note also that, even though this structure presents $O_6$ rotations about a single pseudo-cubic axis, its symmetry is not tetragonal; the reason is that the two directions without rotations (b and c according to the label $a^-b^0c^0$) are obviously symmetry-inequivalent, as one of them is in-plane and the other one out-of-plane. Hence, the structure has an orthorhombic symmetry.

Let us conclude by noting that all the described phases are predicted to be metallic and paramagnetic by our LDA simulations. We inspected the obtained electronic density of states, and ran simulations allowing for spin polarization in selected cases; yet, we did not appreciate any signature that might hint at a transition to a different electronic or magnetic state.

### 6.2 Raman frequencies

We computed the phonon spectra at the Γ q-point for all the phases found, and identified the Raman-active modes. In Fig. 7 we show the frequency of modes in the relevant range of $a_{sub}$ values. While there are significant structural changes across the considered epitaxial strains, it is possible to track the Raman phonons in the relevant frequency window by analyzing the computed eigenvectors. Thus, in Fig. 7 we indicate with dashed lines the approximate evolution of the key phonons as a function of the epitaxial constraint, as identified by comparing the corresponding eigenvectors.

Note that, due to LDA's well-known tendency to overbind, there may be some uncertainty as regards the best strategy to compare our computed Raman frequencies with the experimental data. Here we have fixed $a_{sub}$ to the experimental lattice constants for LAO and LSAT, and thus computed the corresponding Raman spectra. Yet, the LDA-relaxed lattice constants for LAO and LSAT compounds will typically be 1% smaller; thus, had we used the theoretical value for $a_{sub}$, we would have obtained slightly shifted frequencies. Nevertheless, the conclusions of our theory—experiment comparison remain essentially the same irrespective of this choice.

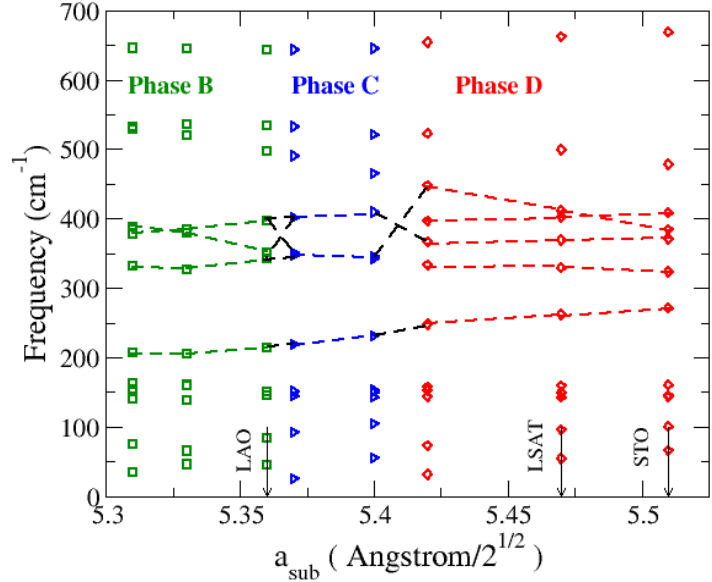

FIG. 7. Computed Raman frequencies for various values of $a_{sub}$. Dashed lines connect phonons whose eigenvectors have significant overlap. The $a_{sub}$ values corresponding to the experimental lattice constants of LAO, LSAT and STO are indicated with vertical arrows.

The comparison between the calculated and experimental Raman frequencies is given in table III. For LNO/LAO, the agreement is remarkably good, which strongly supports our prediction that the "phase B" discussed above is the stable one for the LAO substrate. Note that some of the computed Raman modes do not have an experimental counterpart, which indicates they probably have small intensities associated to them.

For the case of the LSAT substrate, we have a generally good agreement between measured frequencies and our computed values for "phase D". In particular, the splitting discussed above, occurring in the 400 cm$^{-1}$ region when comparing the results for films grown on LAO and LSAT substrates, is clearly appreciated in our calculations as well. In contrast, no splitting of that sort occurs in "phase C", which is the structure that, according to our calculations, becomes stable between phases B and D and the one that May et al.[10] propose to be stable under large tensile strains (e.g., on STO substrates). Further, as mentioned above, the splitting experimentally observed in region III suggests the presence of different Ni—O bond lengths inside the NiO$_6$ octahedra; this is consistent with our predicted "phase D", which is the only one among the calculated structures featuring three different Ni—O lengths (see bottom panel of Fig. 5). Hence, there are good reasons to believe that our newly found "phase D" is indeed the one experimentally obtained, as it permits to explain the observed symmetry breaking.

It is also worth noting that for the LSAT substrate we find one experimental frequency at 218 cm$^{-1}$ that does not have a clear computed counterpart. This frequency corresponds to the soft mode; the apparent disagreement between experiment and theory in this regard is discussed below.

TABLE III. Calculated and experimental frequencies (cm$^{-1}$) for the Raman modes on LNO thin films in various epitaxial conditions. For the theoretical frequencies, we indicate in parenthesis the a$_{sub}$ values at which they were obtained. Bulk values are also included for reference. The soft mode is highlighted in bold.

| Bulk LNO | | Phase B (3.79 Å) | LNO on LAO Exp. | Phase C (3.82 Å) | LNO on LSAT Exp. | Phase D (3.87 Å) |
|---|---|---|---|---|---|---|
| Calc. | Exp. Ref.21 | | | | | |
| --- | --- | --- | --- | --- | --- | --- |
| | | 45 | | 55 | | 54 |
| | | 83 | | 104 | | 95 |
| | | 145 | | 143 | | 142 |
| 70 | 74 | 150 | 156 | 150 | | 149 |
| 162 | 157 | 151 | 167 | | | 159 |
| **217** | **210** | **213** | **214** | **231** | **218** | **261** |
| 388 | 401 | 343 | | 343 | | 329 |
| 528 | 454 | 351 | | 345 | 373 | 368 |
| | | 397 | 403 | 409 | 393 | 402 |
| | | 498 | | 465 | 409 | 412 |
| | | 535 | | 521 | | 499 |
| | | 644 | | 645 | | 662 |

**6.3 Conclusions from the experiment—theory comparison**

The agreement between experiments and simulation results is good at the qualitative and quasi-quantitative levels, and yields a coherent picture of strain-driven structural transitions in LNO films. Yet, there are some aspects in which the interpretation of the experimental results may seem at odds with the theoretical predictions, and which merit some comment.

Most importantly, the magnitude of the tilt angle between oxygen octahedra is particularly relevant to the physics of LNO and nickelates in general, and it is thus pertinent to discuss what our experimental results indicate about its evolution with epitaxial strain. Interestingly, as shown by detailed studies of perovskites in which the tilts of the O$_6$ octahedra are the principal order parameter driving the structural transformations[32–37], the tilt angle is strongly correlated with the frequency of the associated soft mode, which is in turn rather insensitive to other structural features (e.g., bond shortening or lengthening). In the case of LNO, the $A_{1g}$ mode at 214 cm$^{-1}$ has been earlier assigned[20,21] as the antiferrodistortive soft mode of the R-3c structure (this assignment is ratified by our DFT results) and its frequency should allow us to probe changes in the octahedra tilt angle, as the vibration has the pattern of the NiO$_6$ octahedra rotations. More precisely, experimental[21] and theoretical[20] considerations indicate that the position of the $A_{1g}$ mode of LNO should scale by about 23 cm$^{-1}$/deg with the tilt angle. Then, our experimental results in Fig. 4b show that the frequency of the $A_{1g}$ mode in region II undergoes no drastic change with strain: its position varies from 213 cm$^{-1}$ for the most compressively strained 14 nm film on LAO, to 215 cm$^{-1}$ for the unstrained film on silicon, to 218 cm$^{-1}$ for the tensile strained 14 nm film on LSAT. According to the above scaling, this difference of 6 cm$^{-1}$ would correspond to a change in tilt angle of about 0.25°, which is surprisingly modest considering the significant changes in the overall Raman spectrum.

This observation is also in disagreement with the DFT results discussed above and summarized in Figs. 5 and 7. Our simulations show that there is a significant evolution of the tilting pattern in the LNO films, with the rotation axis varying as a function of epitaxial strain (from out-of-plane in the limit of strong in-plane compression, to in-plane for the largest tensions). This structural evolution seems physically sound, and its main features are perfectly in line with what has been predicted to occur in similar rhombohedral perovskites (like $BiFeO_3$[38,39] and others[40]); hence, we do not have any reason to doubt this result. Further, our simulations predict a sizable evolution of the soft-mode frequency with epitaxial strain, from about 215 cm$^{-1}$ for LNO/STO to about 265 cm$^{-1}$ for LNO/LSAT. According to the empirical rule mentioned above, this 50 cm$^{-1}$ shift would suggest a related change of about 2° in the corresponding tilt angle. Naturally, our LDA-predicted structural transition sequence is a complex one, and does not lend itself to a discussion based on a simplified model for the soft-mode frequency. Yet, it is worth noticing that the computed changes in the *total* rotation angle (i.e,. in the modulus of the 3-dimensional rotation vector defined by the rotation amplitudes about the three pseudo-cubic axes) are indeed of the order of 2-3° when moving from STO to LSAT substrates. Hence, in this sense, our LDA results for the soft-mode frequency and tilting magnitude are consistent with expectations from the literature.

All in all, our Raman-measured result for the soft-mode frequency, which is essentially constant for all the measured films, remains surprising. The most likely explanation for this observation is that our films are partly relaxed. Such a partial relaxation can be expected to have a strong influence in some quantitative features of the spectrum – especially on the position of the sensitive soft-mode peak –, while other qualitative features – e.g., peak splittings caused by symmetry breaking – should be less affected by it.

## 7. Concluding remarks

We have presented an experimental and theoretical investigation of compressive and tensile strained $LaNiO_3$ thin films, for which X-ray and TEM analysis attest a high quality. Raman spectroscopy of the thin films provides experimental evidence for structural phase transitions under strain, with different structures under compression and tension. Such multiple phase transitions were hitherto unexpected. Our first-principles calculations provide an in-depth analysis of possible space groups and thus an understanding of the underlying structural mechanisms at play, in terms of octahedral rotations and interatomic distances. Beyond the experimentally observed phase transitions, it is important to note our calculations suggest further instabilities hinting at a rich energy landscape, where several local minima are close in energy. We hope that our study motivates further explorations of this energy landscape by strain, be it through different substrates or the embedding of $LaNiO_3$ in thin film heretrostructures, both currently attracting a significant interest.

**Acknowledgements**

JK, MW, MG and JI acknowledge support from the National Research Fund, Luxembourg through a Pearl Grant (FNR/P12/4853155). ND, FS, GH and JF acknowledge financial support by the Spanish Government (projects SEV-2015-0496, and MAT2014-56063-C2-1-R) and Generalitat de Catalunya (2014 SGR 734).